\documentclass[%
 reprint,
superscriptaddress,
 amsmath,amssymb,
 aps,
]{revtex4-2}

\usepackage{amsmath,amsthm,mathrsfs,amsfonts,dsfont}
\usepackage{graphicx}
\usepackage{color}
\usepackage[10pt]{moresize}
\usepackage{amssymb}
\usepackage{amscd}
\usepackage{enumerate}
\usepackage{epsfig}
\usepackage{subfigure}
\usepackage{bm}
\usepackage{color}
\usepackage{threeparttable}
\usepackage{tabularx}
\usepackage{multirow}
\usepackage {appendix}
\begin{document}

\preprint{APS/123-QED}

\title{Effect of light injection on the security of practical quantum key distribution}% Force line breaks with \\
%\thanks{A footnote to the article title}%

\author{Liying Han}%
\affiliation{Hefei National Research Center for Physical Sciences at the Microscale and School of Physical Sciences, University of Science and Technology of China, Hefei 230026, China}%
\affiliation{Shanghai Research Center for Quantum Science and CAS Center for Excellence in Quantum Information and Quantum Physics, University of Science and Technology of China, Shanghai 201315, China}
\affiliation{Hefei National Laboratory, University of Science and Technology of China, Hefei 230088, China}

\author{Yang Li}
\email{liyang9@ustc.edu.cn}
\affiliation{Hefei National Research Center for Physical Sciences at the Microscale and School of Physical Sciences, University of Science and Technology of China, Hefei 230026, China}%
\affiliation{Shanghai Research Center for Quantum Science and CAS Center for Excellence in Quantum Information and Quantum Physics, University of Science and Technology of China, Shanghai 201315, China}
\affiliation{Hefei National Laboratory, University of Science and Technology of China, Hefei 230088, China}

\author{Hao Tan}
\affiliation{Hefei National Research Center for Physical Sciences at the Microscale and School of Physical Sciences, University of Science and Technology of China, Hefei 230026, China}%
\affiliation{Shanghai Research Center for Quantum Science and CAS Center for Excellence in Quantum Information and Quantum Physics, University of Science and Technology of China, Shanghai 201315, China}
\affiliation{Hefei National Laboratory, University of Science and Technology of China, Hefei 230088, China}

\author{Weiyang Zhang}
\affiliation{Hefei National Research Center for Physical Sciences at the Microscale and School of Physical Sciences, University of Science and Technology of China, Hefei 230026, China}%
\affiliation{Shanghai Research Center for Quantum Science and CAS Center for Excellence in Quantum Information and Quantum Physics, University of Science and Technology of China, Shanghai 201315, China}
\affiliation{Hefei National Laboratory, University of Science and Technology of China, Hefei 230088, China}

\author{Wenqi Cai}
\affiliation{Hefei National Research Center for Physical Sciences at the Microscale and School of Physical Sciences, University of Science and Technology of China, Hefei 230026, China}%
\affiliation{Shanghai Research Center for Quantum Science and CAS Center for Excellence in Quantum Information and Quantum Physics, University of Science and Technology of China, Shanghai 201315, China}
\affiliation{Hefei National Laboratory, University of Science and Technology of China, Hefei 230088, China}

\author{Juan Yin}
\affiliation{Hefei National Research Center for Physical Sciences at the Microscale and School of Physical Sciences, University of Science and Technology of China, Hefei 230026, China}%
\affiliation{Shanghai Research Center for Quantum Science and CAS Center for Excellence in Quantum Information and Quantum Physics, University of Science and Technology of China, Shanghai 201315, China}
\affiliation{Hefei National Laboratory, University of Science and Technology of China, Hefei 230088, China}

\author{Jigang Ren}
\affiliation{Hefei National Research Center for Physical Sciences at the Microscale and School of Physical Sciences, University of Science and Technology of China, Hefei 230026, China}%
\affiliation{Shanghai Research Center for Quantum Science and CAS Center for Excellence in Quantum Information and Quantum Physics, University of Science and Technology of China, Shanghai 201315, China}
\affiliation{Hefei National Laboratory, University of Science and Technology of China, Hefei 230088, China}

\author{Feihu Xu}
\affiliation{Hefei National Research Center for Physical Sciences at the Microscale and School of Physical Sciences, University of Science and Technology of China, Hefei 230026, China}%
\affiliation{Shanghai Research Center for Quantum Science and CAS Center for Excellence in Quantum Information and Quantum Physics, University of Science and Technology of China, Shanghai 201315, China}
\affiliation{Hefei National Laboratory, University of Science and Technology of China, Hefei 230088, China}

\author{Shengkai Liao}
\email{skliao@ustc.edu.cn}
\affiliation{Hefei National Research Center for Physical Sciences at the Microscale and School of Physical Sciences, University of Science and Technology of China, Hefei 230026, China}%
\affiliation{Shanghai Research Center for Quantum Science and CAS Center for Excellence in Quantum Information and Quantum Physics, University of Science and Technology of China, Shanghai 201315, China}
\affiliation{Hefei National Laboratory, University of Science and Technology of China, Hefei 230088, China}

\author{Chengzhi Peng}
\affiliation{Hefei National Research Center for Physical Sciences at the Microscale and School of Physical Sciences, University of Science and Technology of China, Hefei 230026, China}%
\affiliation{Shanghai Research Center for Quantum Science and CAS Center for Excellence in Quantum Information and Quantum Physics, University of Science and Technology of China, Shanghai 201315, China}
\affiliation{Hefei National Laboratory, University of Science and Technology of China, Hefei 230088, China}

\date{\today}% It is always \today, today,
             %  but any date may be explicitly specified

\begin{abstract}
Quantum key distribution (QKD) based on the fundamental laws of quantum physics can allow the distribution of secure keys between distant users. 
However, imperfections in realistic devices may lead to potential security risks, which must be accurately characterized and considered in practical security analysis.
High-speed optical modulators, being one of the core components of practical QKD systems, can be used to prepare the required quantum states. 
Here, we find that optical modulators based on $LiNbO_3$, including phase modulators and intensity modulators, are vulnerable to photorefractive effects caused by external light injection. 
By changing the power of external light, eavesdroppers can control the intensities of the prepared states, posing a potential threat to the security of QKD.
We have experimentally demonstrated the influence of light injection on $LiNbO_3$-based optical modulators and analyzed the security risks caused by the potential light injection attack, along with the corresponding countermeasures.
%\begin{description}
%\item[Usage]
%Secondary publications and information retrieval purposes.
%\item[Structure]
%You may use the \texttt{description} environment to structure your abstract;
%use the optional argument of the \verb+\item+ command to give the category of each item. 
%\end{description}
\end{abstract}

%\keywords{Suggested keywords}%Use showkeys class option if keyword
                              %display desired
\maketitle

%\tableofcontents

\section{\label{sec:level1}Introduction}
Since the first protocol proposed in 1984 \cite{Bennett1984}, quantum key distribution (QKD), which can enable two distant parties Alice and Bob to share the same secret keys with information-theoretical-proven security, has made tremendous developments in the past few decades \cite{Bennett1989ACS,Peng2007RPL,rosenberg2007longPRL,Liu2023PRL,Buttler1998PRL,Ursin2007NP,Liao2017NP,yin2020,chen2021,RN73}. 
However, in practical QKD systems, there is a gap between the devices and the theoretical model, and imperfect devices could be exploited by an eavesdropper to obtain information about the secret keys \cite{Xu2020RMP,THA1,THA2,THA3,sun2015,pang2020,huang2019,huang2020,Ponosova2022,Xu_2010,Wu:20,qi2007,Vadim2011,Tan_2022}. 
Many methods have been proposed to defeat the attacks. 
Security patching is a practical approach, that is, once one discovers a new type of attack, corresponding countermeasures against this attack can be proposed and realized in the existing QKD systems. The second approach is to fully characterize the devices used in a QKD system and describe the devices accurately in mathematical models.  
Moreover, measurement-device-independent QKD (MDI-QKD) \cite{lo2012} and twin-field QKD (TF-QKD) \cite{TF2018} have been recently proposed, which are immune to all the potential detection-side loopholes. 
The source side becomes the more vulnerable part of any QKD setup and attracts more attention.
% Attacks at practical QKD systems include Trojan horse attack \cite{THA1,THA2,THA3}, laser seeding attack \cite{sun2015,pang2020,huang2019}, laser-damage attack \cite{huang2020,Ponosova2022},phase-remapping attack \cite{Xu_2010}, detector blinding attack \cite{Wu:20}, time-shift attack \cite{qi2007}, channel calibration attack \cite{Vadim2011}, etc.

There have been several researches on the potential loopholes at the source-side components of practical QKD systems \cite{THA1,THA2,sun2015,pang2020,huang2019,huang2020,Ponosova2022}, including laser diodes (LDs), attenuators, optical modulators, isolators and so forth. 
By injecting intense external light into the LDs, the phase, wavelength or intensity of the output light may be changed and some side-channel information can be extracted by the eavesdroppers \cite{sun2015,pang2020,huang2019}. 
There are also some attacks against optical modulators, such as Trojan horse attack \cite{THA1,THA2,THA3} and phase-remapping attack \cite{Xu_2010}, where phase information can be extracted from a phase modulator (PM) using optical-frequency-domain reflectometry or from the imperfection of electrical pulses for modulation. 
Meanwhile, the recent work demonstrates that it is not foolproof to adopt optical isolators and attenuators to resist external light \cite{Tan_2022}, as their performance deteriorates in the presence of an external magnetic field and external strong light. 
% There is no doubt that optical modulators are key device of QKD systems, so research on its attack is very significant to the security of practical QKD.

Currently, high-speed optical modulators are widely applied in practical QKD systems to prepare the required quantum states for different protocols, and $LiNbO_3$-based optical modulators are the most  used type due to the wide transparent window, high refractive index, high second-order nonlinearity and stable physical and chemical characteristics \cite{QiLi2020}.
Lithium niobate is a kind of photorefractive material, whose refractive index distribution can be controlled by external illumination \cite{ma2012}. And the light-induced $\Delta$n produces beam degradation during propagation and light intensity limitation effects \cite{Jubera2014,Villarroel:10}. 
Based on this characteristic, we propose and demonstrate a method of using external green light injection to affect the security of QKD system. 
Our experiment demonstrates that all the tested modulators show increased insertion loss after green light irradiation, with a maximum loss change of 19.53 dB for the PM samples and 1.31 dB for the intensity modulator (IM) samples.
These insertion loss variations can be further recovered and restored to the initial state by shining weaker green light on the modulator.
Taking advantage of this ability to proactively control the intensity of the emitted quantum light, the security of practical QKD systems may be compromised by eavesdroppers.
Based on our experimental results, we then analyze the security risks caused by the potential green light injection attack and discuss about the corresponding countermeasures.

This paper is organized as follows. In section \ref{LiNbO_3}, we give an introduction to $LiNbO_3$-based modulators in QKD systems. In section \ref{exp_setup}, we show the experimental setup of green light injection and the  results with several modulator samples. In section \ref{Security}, we analyze the effect of this phenomenon on the security of QKD. In section \ref{Countermeasures}, we discuss some potential countermeasures against the hacking strategy proposed. Finally, we make conclusions in section \ref{Conclusion}.

\section{\label{LiNbO_3}$LiNbO_3$-based modulators in quantum key distribution}
\begin{figure*}
\centering
\includegraphics[width=0.9\linewidth]{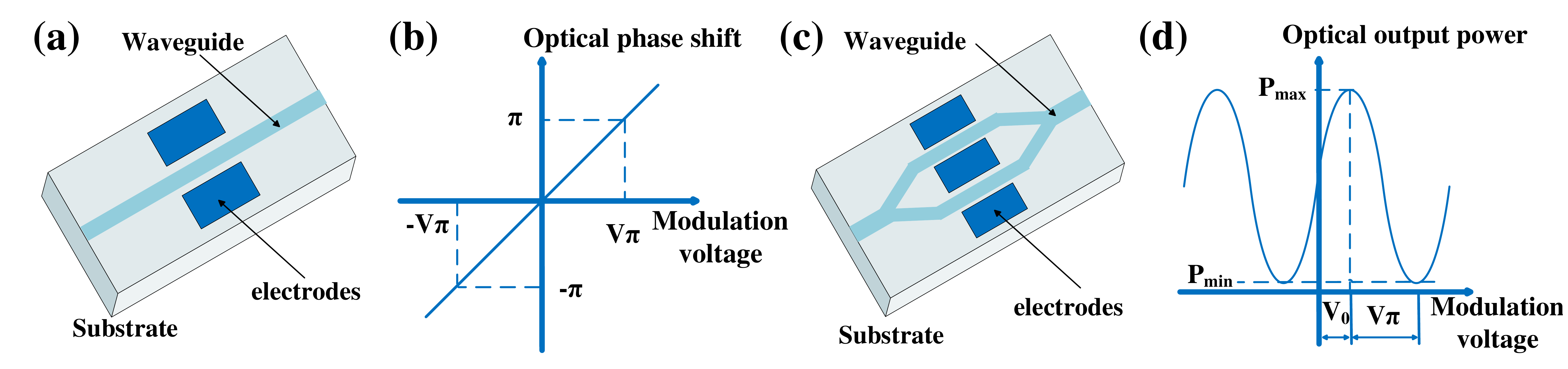}
\caption{
(a) Internal structure and components of phase modulators, including a substrate, a $LiNbO_3$-based waveguide and electrodes that provide an external electric field. 
(b) Phase modulator characteristic curve.
(c) Internal structure and components of intensity modulators, which uses a Mach–Zehnder interferometer configuration with two branches. 
(d) Intensity modulator characteristic curve. 
The applied voltage will cause a phase difference between the two branches, resulting in a change in output power.
}
\label{fig1}
\end{figure*}

$LiNbO_3$-based optical modulators take advantage of linear electro-optic effect to realize optical phase shift, which is an optically second-order non-linear effect, also known as the Pockels-Effect \cite{Liu2015}. 
This effect describes the change in the refractive index of an optical material when an external electric field is applied, and the change in refractive index is directly proportional to the applied electric field.
The internal structure and components of phase modulators and intensity modulators are depicted in Fig. \ref{fig1}(a) and Fig. \ref{fig1}(c), respectively.
Furthermore, Fig. \ref{fig1}(b) and Fig. \ref{fig1}(d) display the corresponding modulator characteristic curves.

% The amount of change in refractive index is proportional to the strength of the electric field, as described in Fig. \ref{fig1}(b).
Based on the characteristic of optical phase shift, $LiNbO_3$-based optical modulators are widely applied in QKD applications, including both decoy state encoding \cite{wang2005,lo2005} and quantum state encoding.
The decoy state encoding can be realized either with a direct $LiNbO_3$ IM \cite{Xu2020RMP}, or with an intensity modulator built using a Sagnac interferometer and a $LiNbO_3$ PM \cite{Roberts:2018}.
The quantum state encoding $LiNbO_3$ using phase modulators includes phase encoding \cite{PhysRevA.101.032319}, time-bin phase encoding \cite{PhysRevLett.121.190502}, polarization encoding \cite{Li:19} and Gaussian discrete modulations \cite{ma2014}, etc. 
% \textcolor{red}{Internal structures of PM and IM are shown in \ref{fig1}(a) and (c), and modulation effect is shown in \ref{fig1}(b) and (d).}

Photorefractive is another optical non-linear effect of $LiNbO_3$, which results from the photovoltaic effect caused by external illumination \cite{kosters2009,zhang1996}. 
The generation process of the photorefractive effect is as follows: external light drives photoexcited charges into adjacent bands and forms the photogenerated charge carriers; the charges leave the light area and settle in the dark area driven by the carrier concentration distribution, extra electric field or photovoltaic effect thus; the Space-charge distribution induces a space-charge field, resulting in a change in the refractive index distribution. 
For $LiNbO_3$ crystals, the photovoltaic effect is the main mechanism for the movement of photoexcited carriers in the absence of an applied electric field.
Under the exposure of the photoelectric field, the refractive index distribution of the waveguide undergoes a permanent change.

The photorefractive effect, as a nonlinear optical phenomenon, exhibits distinct behaviors depending on the intensity of the incident light \cite{Villarroel:10, Jubera2014}. 
In the case of weak light injection, the photorefractive effect is primarily characterized by a phase variation, leading to a phase shift in the transmitted or modulated light. 
However, when subjected to strong light injection, the photorefractive effect becomes more pronounced, giving rise to significant alterations in the spatial profile, intensity, and direction of the light beam as it traverses the $LiNbO_3$ crystals. 
This strong effect can induce phenomena such as diffraction, self-focusing, or beam steering, resulting in beam distortion and subsequently causing an increase in loss for the $LiNbO_3$ modulators.

% This change of refractive index will not disappear immediately in the absence of external light, can be preserved for quite a long time - days or even months - that is why $LiNbO_3$ can be used for data storage. 

The photorefractive effects observed in $LiNbO_3$ modulators are distinct due to their susceptibility to relatively low light powers compared to other nonlinear phenomena. 
In particular, the induction of photorefractive phenomena requires microwatt-level light, while milliwatt-level light can lead to beam distortion and alteration of the modulators' loss characteristics. 
Importantly, the change of refractive index will not disappear immediately in the absence of external light.
In the case of normal placement and without additional measures, the recovery of the photorefractive effect in the crystal will take quite a long time - days or even months - that is why $LiNbO_3$ can be used for data storage. 
To expedite the recovery process, various measures can be employed, including uniform illumination or thermal treatments \cite{Jubera2014, zhang1996}. 
Exploiting the unique characteristics of photorefractive, an eavesdropper can potentially manipulate the loss of $LiNbO_3$ modulators by strategically utilizing external light sources. 
This deliberate control can compromise the security of practical QKD systems.

\section{\label{exp_setup}Experimental measurements on light injection}
\subsection{Experimental setup}
In our experiments, we select several $LiNbO_3$ modulator samples to test their properties before and after green light irradiation, the basic information of which is summarized in Table \ref{tab1}. 
The parameters to be tested include insertion loss, half-wave voltage (V$\pi$), and extinction ratio (only for IM).
For PM, the insertion loss is the ratio of the output power to the input power in the absence of impressed voltage, and the V$\pi$ is the voltage required to increase the phase by $\pi$.
For IM, the insertion loss is the ratio of the output power to the input power, where the output power is defined as the sum of maximum transmitted power and minimum transmitted power with impressed voltage in a half-wave voltage cycle. 
V$\pi$ is the voltage required to change the output power from minimum to maximum, and the extinction ratio is the result of dividing the maximum and minimum power.
% As for PMs, insertion loss and half-wave voltage (V$\pi$) are main parameters to be concerned in use. 
% We focus on testing their changes and whether the changes can be restored or not. 
% As for IMs, the extinction ratio is also tested except for insertion loss and V$\pi$. 

\begin{table*}[ht!]
\renewcommand{\arraystretch}{1.5}
\caption{Basic information on all the modulators under test (5 phase modulators and 2 intensity modulators), including manufacturer, waveguide process, Doping process and Model. 
Among them, PM-1, PM-2 and PM-5 are from the same manufacturer Conquer, while PM-1 and PM-5 belong to the same batch.
$LiNbO_3$-based modulators are usually fabricated using two processes, Ti diffusion or proton exchange, along with a selective doping process to increase the threshold of photorefractive resistance. 
The term "Unspecified" indicates that no information was obtained from the manufacturer.
}
\label{tab1}
\begin{tabular}{ccccc}
\hline
Number&Manufacturer&Waveguide process&Doping process&Model
\\\hline
PM-1&Conquer&Ti diffusion&Undoped&KG-PM-15-10G-PP-FA\cr
PM-2&Conquer&Ti diffusion&Undoped&KG-PM-15-10G-PP-FA-H\cr
PM-3&Ixblue&Unspecified&MgO&MPZ-LN-10\cr
PM-4&Eospace&Unspecified&Unspecified&PM-5S5-10-PFAPFA-UV\cr
PM-5&Conquer&Ti diffusion&Undoped&KG-PM-15-10G-PP-FA\cr
IM-1&Conquer&Proton exchange&Undoped&KG-AM-HER-10G-PP-FA\cr
IM-2&Eospace&Unspecified&Unspecified&AZ-0S5-10-PFAPFA\cr 
\hline
\end{tabular}
\end{table*}

\begin{figure*}
\centering 
\includegraphics[width=1\linewidth]{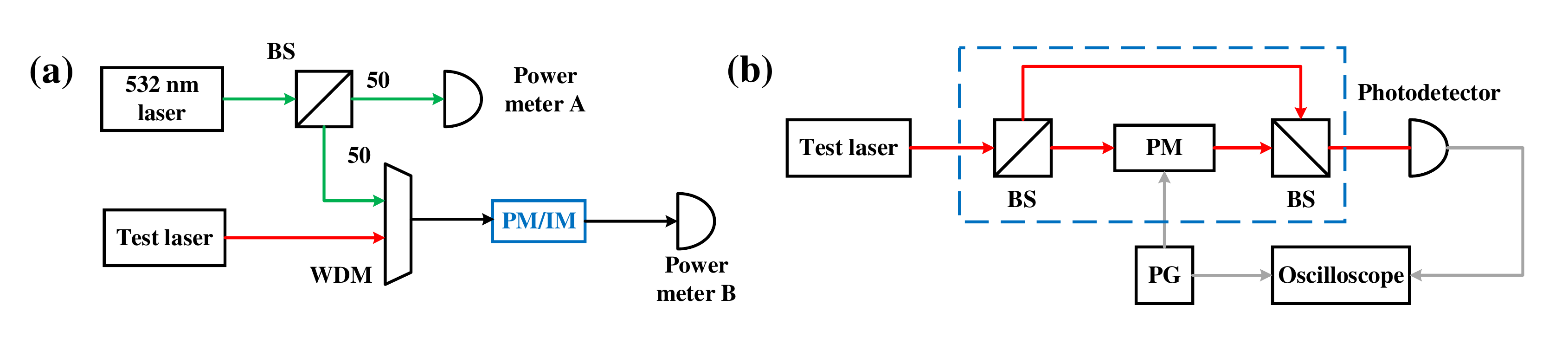}
\caption{Simplified diagram of the experimental setup of (a) insertion loss, with the PM/IM as a replaceable device under test.
BS: beam splitter; WDM: wavelength division multiplexer.
(b) V$\pi$, with the PM in an MZI configuration or IM as a replaceable device under test.
PG: electronic pulse generator; MZI: Mach-Zehnder interferometer.
}
\label{fig2}
\end{figure*}

The test of insertion loss and the V$\pi$ of IM is performed using the setup shown in Fig. \ref{fig2}(a), which is an all-fiber setup.
The test laser is a fiber-pigtailed 1550 nm LD, combined with the 532 nm injection laser (Cnilaser MGL-III-532-50mW) using a wavelength division multiplexer (WDM).
Power meter A (Thorlabs PM400) monitors the power of the 532 nm laser after a 50:50 fiber beam-splitter (BS), while power meter B is used to check the loss of modulator before and after green light injection. 
% \textcolor{red}{As for PMs, the insertion loss is the ratio of the output power to the input power in the absence of impressed voltage. 
% For IMs, we apply and scan the DC voltage to obtain  maximum and minimum transmitted power, whose sum is defined as insertion loss.}
In order to test V$\pi$ of PM, we use the PM to construct a Mach-Zehnder-interferometer configuration, which can be replaced when testing the IMs, as shown in Fig. \ref{fig2}(b). 
The electronic pulse generator (Tektronix AFG3022C) produces two sawtooth waves of the same frequency, one for triggering the oscilloscope (WaveRunner 8254M) and the other for modulating the PM/IM. 
The electrical signal output by the photodetector (Newport 1544-B) is also connected to the oscilloscope, similar ot the curve in Fig. \ref{fig1}(d). 
V$\pi$ can be obtained by calculating the period of the measured sinusoidal curve. 
The extinction ratio of the IM can be obtained by applying accretive DC voltage and measuring the maximum and minimum power within a half-wave voltage cycle.

\subsection{Experimental results}
\subsubsection{Phase modulator}
\begin{figure*}
\centering 
\includegraphics[width=0.9\linewidth]{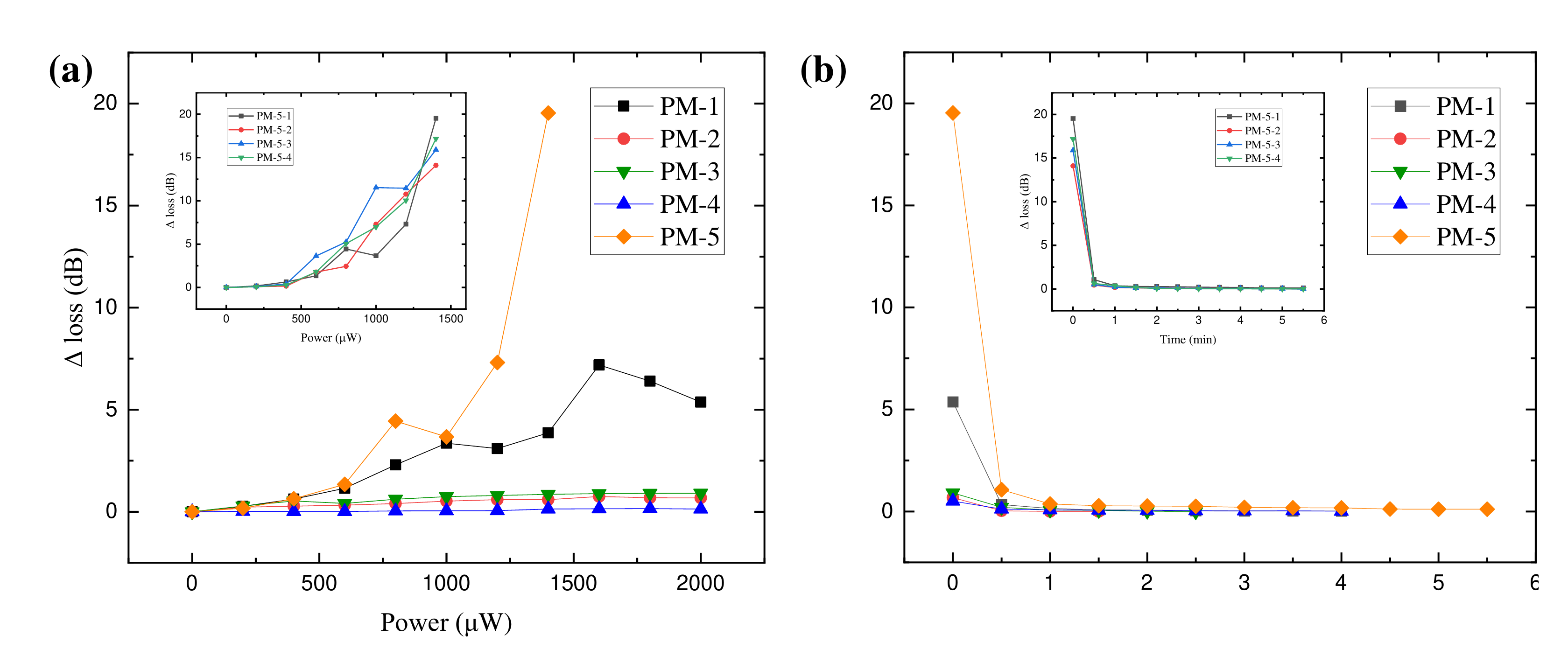}
\caption{(a) Test results of increased insertion loss of PMs. 
We increase the power of the incident green light to 2 mW with a step size of 200 $\mu$W and an exposure time of 5 minutes per step.
The PMs show an increase in insertion loss with a maximum of 19.53 dB.
(b)  Test results of recovered insertion loss of five PMs.
We inject 50 $\mu$W of green light to restore the insertion loss of modulators with an exposure time of 30 seconds per step.
All the PMs rapidly restore insertion loss to original state within 5 minutes.
The subplots in both figures depict the results of four repeated tests conducted on PM-5. 
Although there are slight variations observed among the four tests, the overall trend of the test results remains consistent.
}
\label{fig3}
\end{figure*}

We first test the insertion loss of five PM samples before and after green light irradiation, and the test results are shown in Fig. \ref{fig3}. 
We increase the power of the incident green light to 2 mW with a step size of 200 $\mu$W and an exposure time of 5 minutes per step.
All five PMs show an increase in insertion loss with increasing green light power.
Particularly, PM-1 and PM-5 exhibit maximum insertion loss increases of about 7.19 dB and 19.53 dB, while PM-2 and PM-3 exhibit similar performance with insertion loss increases of slightly less than 1 dB. 
At the same time, PM-4 does not seem to be greatly affected by the green light of 2 mW, and its insertion loss increases by 0.5 dB as the green light power continues increase to 8 mW. 
We will not continue to increase the optical power in case irreversible photo damage occurs.
To recover the insertion loss, we inject 50 uW of green light into the PMs with an exposure time of 30 seconds per step and record the insertion loss over time. 
All five PMs quickly recover their insertion loss within 4 minutes within the error range.
It should be noted that the alteration illumination and the recovery illumination are indistinguishable apart from the difference in light intensity. 
No additional specialized procedures were performed during the recovery phase.

We also test the V$\pi$ of five PMs before, after green light irradiation, and after recovery, and the test results are shown in Table \ref{tab2}. 
Particularly, the half wave voltages of PM-1 and PM-5 show maximum increases of 1.53 V and 2.81 V after green light irradiation, and can also be recovered to their original state within the error range.

\begin{table}[ht!]
\renewcommand{\arraystretch}{1.5}
\caption{Test results of half wave voltage of PMs before green light irradiation (V$\pi_{B}$), after green light irradiation (V$\pi_{A}$) and after recovery (V$\pi_{R}$). 
$\Delta$V$\pi$ is the increase of V$\pi$ after green light irradiation. 
$\Delta$Loss is the maximum increase of insertion loss in the process of green light irradiation. The half wave voltage of the PMs shows a maximum increase of 2.81 V after green light irradiation, and can also be recovered to its original state.}
\label{tab2}
\begin{tabular}{cccccc}
\hline                             
Number&V$\pi_{B}$ (V)&V$\pi_{A}$ (V)&V$\pi_{R}$ (V)&$\Delta$V$\pi$ (V)&$\Delta$Loss (dB)\cr 
\hline  
PM-1&4.04&5.57&4.03&1.53&7.19\cr
PM-2&3.90&4.06&3.91&0.16&0.75\cr
PM-3&4.68&5.10&4.70&0.42&0.91\cr
PM-4&2.79&2.79&2.78&0.00&0.50\cr
PM-5&3.99&6.80&4.00&2.81&19.53\cr
\hline  
\end{tabular}
\end{table}

From the above test results, it can be seen that the five PMs mainly exhibit the phenomenon of increased insertion loss and fast recovery under weaker green light injection, which is obviously consistent with the photorefractive characteristics.
Despite being from the same manufacturer, PM-2 shows better performance than PM-1 and PM-5, indicating variability between batches. 
% \textcolor{red}{We also test another PM, PM-5, from the same batch with PM-1,and the results shows that the maximum loss increased by about 20 dB and V$\pi$ increased by 2.81 V after green light irradiation.}
Therefore, we suspect that the batch of lithium niobate crystals used for PM-2 is of better quality and has fewer defects, resulting in less variation of insertion loss. 
As for PM-3, we learned from the manufacturer that PM is doped with an optical-damage-resistant Impurities-MgO, which can effectively increase the optical damage threshold \cite{lengyel2015growth} and may explain the smaller variation of insertion loss than PM-1. 
Unfortunately, we did not get useful information about PM-4 from its manufacturer, such as the doping process and waveguide process.
Therefore, we cannot draw any conclusion about this. 

% There are other recovery methods, such as placing the modulator in the dark and heating it \cite{zhang1996}. 

After irradiation with 2 mW green light, we placed the PM-1 normally without additional measures for 3 days, and the insertion loss recovered about 1.56 dB without additional irradiation. 
The loss recovery values of the other four modulators are 0.08 dB, 0.17 dB, 0.11 dB and 3.50 dB, respectively.
Noting that the insertion loss does not restore to its initial value after 3 days of normal storage, and it will take longer to fully recover. 
This works in Eve's favor because the injection green laser doesn't need to be turned on all the time, making Eve less likely to be found.

% The other phenomenon is that half wave voltage of PMs shows increase correlated with the variation of insertion loss. 
% Photorefractive causes change of the refractive index distribution, that can lead to the increase of half wave voltage. 
% The variation of insertion loss reflects the degree of refractive index change from the side, so there is positive correlation between half wave voltage and insertion loss.

\subsubsection{Intensity modulator}
\begin{figure*}
\centering 
\includegraphics[width=0.9\linewidth]{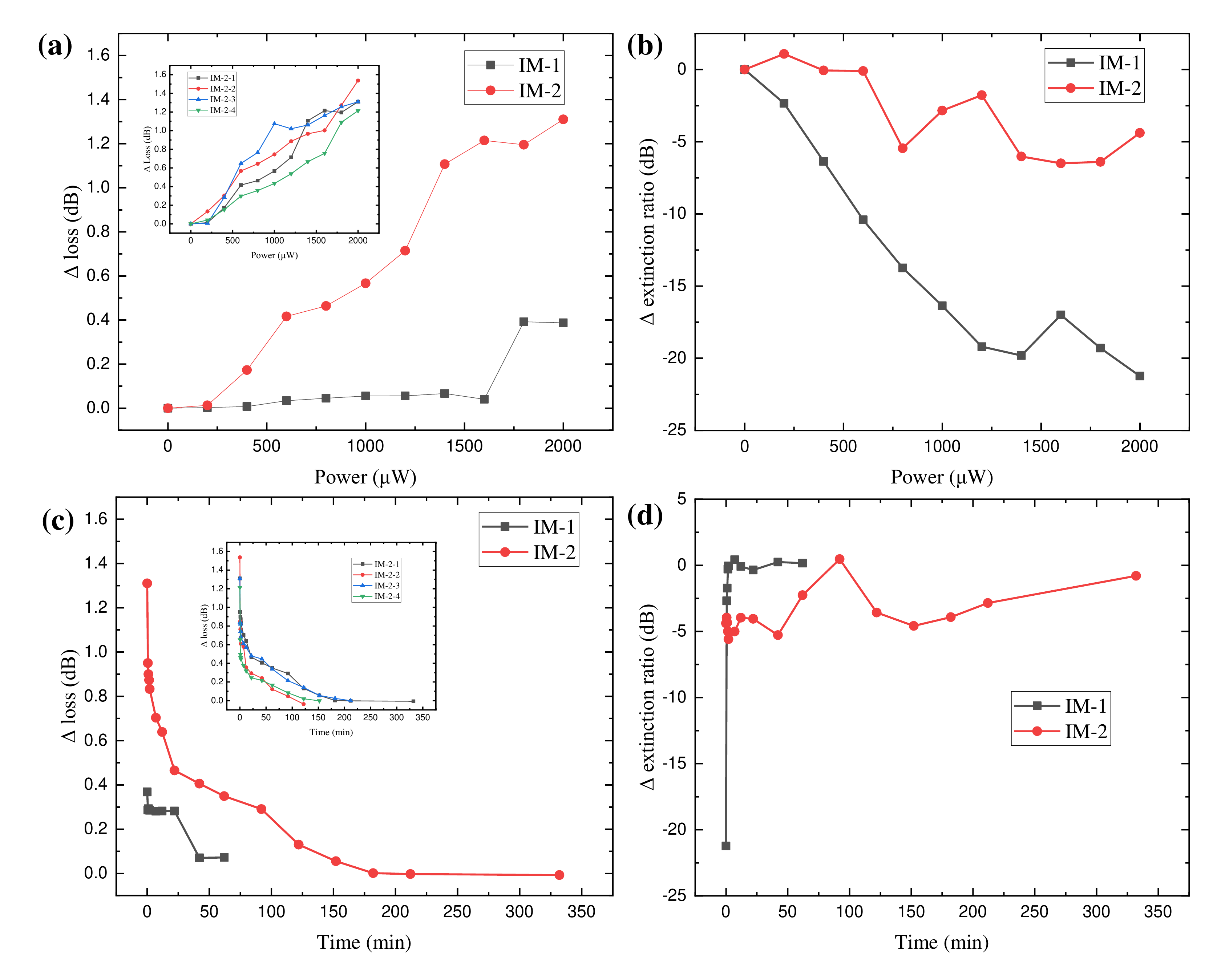}
\caption{(a) Test results of increased insertion loss of two IMs. 
We increase the power of the incident green light to 2 mW with a step size of 200 $\mu$W and an exposure time of 5 minutes per step.
The IMs show an increase in insertion loss with a maximum of 1.31 dB.
(b) Test results of decreased extinction ratio of two IMs.
The IMs show a decrease in extinction ratio with a maximum of 21.23 dB.
(c) Test results of recovered insertion loss of two IMs.
All two IMs restore insertion loss to the original state within 330 minutes.
(d) Test results of recovered extinction ratio of two IMs.
IM-1 restores insertion loss to the original state within 60 minutes, while IM-2 does not restore to the original state after 330 minutes.
The subplots in subfigure a and subfigure c depict the results of four repeated tests conducted on IM-2. 
Despite the minor variations among the four tests, the overall trend of the test results remains consistent.}
\label{fig4}
\end{figure*}

The test results of insertion loss and extinction ratio of two IM samples are shown in Fig. \ref{fig4}. 
Similar to PMs, the insertion loss of IMs generally increases with the power of injection green light, with a insertion loss increase of about 0.39 dB for IM-1 and 1.31 dB for IM-2. 
As for the recovery process, it takes about 60 minutes for IM-1 and about 330 minutes for IM-2 to restore to their original states, which are much longer than the above PMs.
The extinction ratio of IMs also decreases as the power of injection green light increase, where IM-1 decreases from 44.39 dB to 23.16 dB and IM-2 decreases from 24.27 dB to 17.77 dB.
The decrease of extinction ratio can also be recovered with the injection of 50 $\mu$W of green light.
The test results of the half-wave voltage of two IMs are shown in Table \ref{tab3}, and the maximum change is less than 0.1 V.
% The above tests demonstrate that insertion loss of $LiNbO_3$-based IMs can also be controlled by external green light, but may take more time than above PMs.
% Compared with the recovery process of PMs, the time consumed by IMs is much longer. 
% However, all the tests above can still demonstrate that insertion loss of $LiNbO_3$-based IMs can be controlled by external green light. 
% test results of decreased  increases in general, and returns to the initial values after the recovery process. 

\begin{table}[ht!]
\renewcommand{\arraystretch}{1.5}
\caption{Half wave voltage of IMs before green light irradiation (V$\pi_{B}$), after green light irradiation (V$\pi_{A}$) and after recovery (V$\pi_{R}$). 
$\Delta$V$\pi$ is the increase of V$\pi$ after green light irradiation.
$\Delta$Loss is the maximum increase of insertion loss in the process of green light irradiation.}
\label{tab3}
\begin{tabular}{cccccc}
\hline                             
Number&V$\pi_{B}$ (V)&V$\pi_{A}$ (V)&V$\pi_{R}$ (V)&$\Delta$V$\pi$ (V)&$\Delta$Loss (dB)\cr 
\hline 
IM-1&5.00&5.06&5.06&0.06&0.39\cr
IM-2&4.04&4.11&4.04&0.07&1.31\cr
\hline  
\end{tabular}
\end{table}

From the above test results, IM-1 outperforms IM-2 with less increase in insertion loss. 
We learned from the manufacturer that PM-1 uses a waveguide process of Ti diffusion, while IM-1 uses a waveguide process of proton exchange, and its photorefractive sensitivity is much lower than that of Ti-indiffused waveguide \cite{Kondo:94}. 
At the same time, we do not have process information from which to draw conclusions explaining the poor performance of IM-2. 
Compared with the PMs, IM exhibits different phenomena in terms of extinction ratio decrease.
Here, the splitting ratio of the 1550 nm Mach–Zehnder modulator is not 50 : 50 at 532 nm, which results in a different insertion loss increase of the two arms.
This will in turn causes the splitting ratio to be no longer 50:50 and result in a decreasing extinction ratio.
% Compared with the PMs, IM exhibit different phenomenons in terms of extinction ratio decrease and longer time taken to cover insertion loss. 
% \textcolor{red}{Meanwhile, the recovery rate of the attenuation of the two arms is different, that results from the different power of 532 nm light divided into two arms to restore the insertion loss. 
% Thus, the time consumed by IMs is much longer than PMs.}

\section{\label{Security}Security risk evaluation of light injection attack}
\begin{figure}
\centering 
\includegraphics[width=1\linewidth]{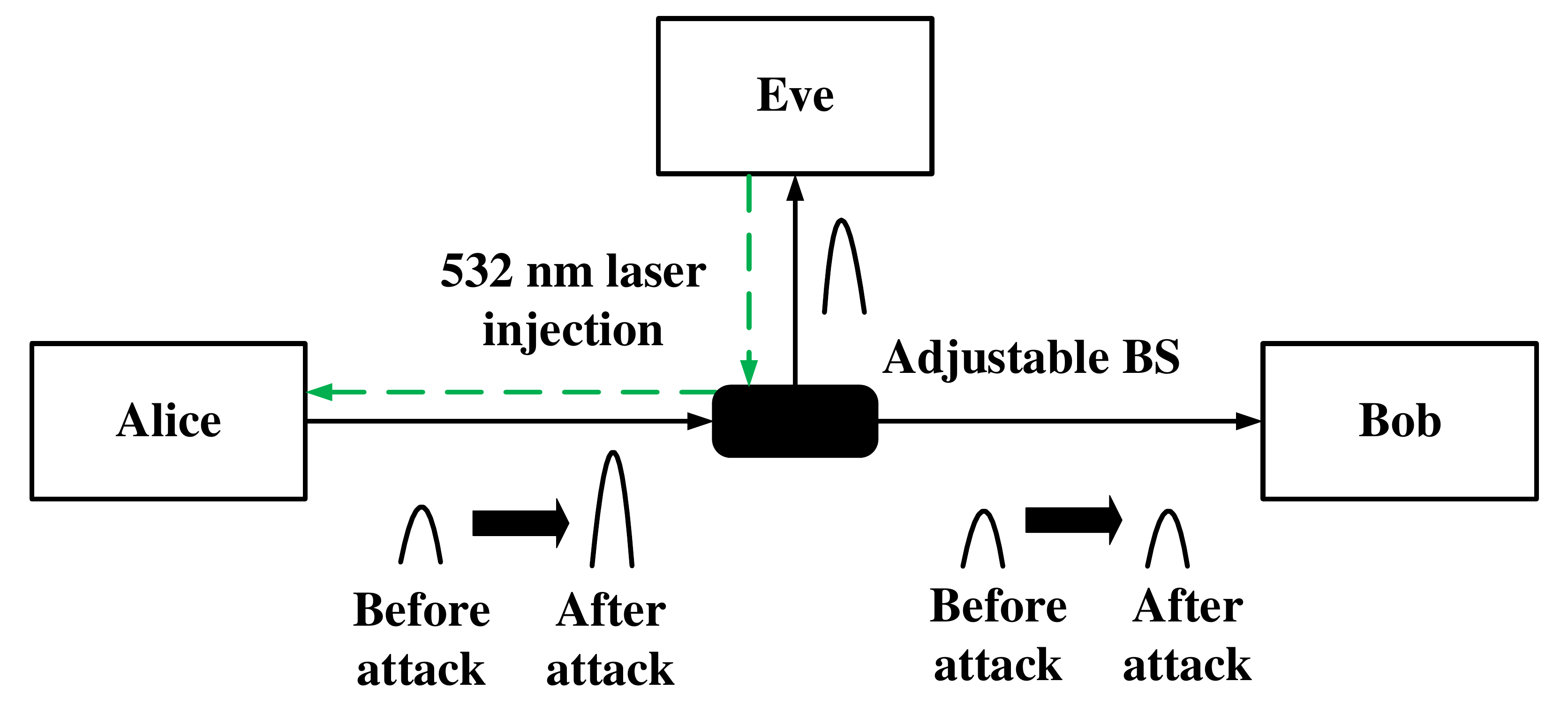}
\caption{Basic schematic of the green light injection attack.
An eavesdropper can actively control the output photon numbers of the QKD transmitter through external green light injection, while ensuring that Bob receives an unchanged number of photons. 
}
\label{fig5}
\end{figure}

The basic schematic of the green light injection attack is shown in Fig. \ref{fig5}.
We put no assumption on the quantum channel used for QKD, and the channel is completely open to eavesdroppers. 
The eavesdropper, Eve, can launch a green light injection attack in three steps: 
\begin{enumerate}[(1)]
\item Eve injects strong green light from the quantum channel into the QKD transmitter (Alice) to increase the insertion loss of modulators without the owner noticing.
\item The owner of the QKD system calibrates the QKD system, which means Alice and QKD receiver (Bob) accept the fact that the calibrated QKD system is in the initial state and the owners start quantum communication.
\item Eve injects weaker green light to restore the insertion loss of modulators.
Meanwhile, Eve can utilize a BS with an adjustable beam splitting ratio to acquire a copy of the quantum light. 
By monitoring the number of detected photons in real time, Eve can assess the changes in loss at Alice's modulators and adjust the light splitting ratio of the BS accordingly. 
Throughout this process, Eve can ensure that Bob receives an unchanged number of photons. 
As a result, Eve and Bob possess identical quantum states and decoding mechanisms. 
After performing the measurement, Eve can obtain the corresponding key, compromising the security of the system.

\end{enumerate}
% Firstly, Eve injects strong green light from quantum channel into the QKD transmitter (Alice) to increase the insertion loss of modulators without the owner noticing; secondly, the owner of the QKD system calibrates the QKD system, which means Alice and QKD receiver (Bob) accept the fact that the calibrated QKD system is in the initial state and the owners start quantum communication; thirdly, Eve injects weaker green light to restore the insertion loss of modulators. 
% By repeating the first and third steps to control insertion loss of modulators, Eve can actively control the output photon numbers of the QKD transmitter without drawing attention.
\textcolor{red}{
% After that, Eve can use a BS with adjustable beam splitting ratio to get copy of the qubits while keeping the parameters of legitimate users unchanged. At this point, Eve and Bob have the exact same quantum states and decoding device. After measurement, Eve can obtain the corresponding keys.
}

In the proposed attack, the process of alteration illumination and recovery illumination each takes approximately 5 minutes. 
This attack is significantly shorter in duration compared to the days-long recovery time required for the modulator insertion loss to return to its natural state. 
By taking advantage of this short attack window, it becomes possible to perform offline attacks, thereby greatly reducing the probability of the eavesdropper being detected.
% As the experimental results shown in section 3, the output power of QKD transmitter becomes controllable for Eve after the green light injection. 

We adopt the security analysis of QKD in Ref. \cite{huang2019} to the prepare-and-measure decoy-state BB84 protocol. 
The typical implementation is evaluated where Alice and Bob use three different intensities, $\mu_s$, $\nu_1$, and $\nu_2$ that satisfy $\mu_s > \nu_1 > \nu_2$ and $\nu_2 = 0$.
Secret keys can be extracted from those events employing the signal intensity $\mu_s$ in the Z basis and X basis, while the decoy $\nu_1$ intensity events are used for parameter estimation. 
In the asymptotic limit of an infinite number of transmitted signals, the secret key rate can be lower bounded by \cite{ma2005}
\begin{equation}
    R_L\geq\frac{1}{2}(-Q_\mu f_eH_2(E_\mu)+Q_1(1-H_2(e_1))),
\end{equation}
where Q$_\mu$ is the gain  of signal states, E$_\mu$ is the overall quantum bit error rate, Q$_1$ is the gain of single-photon states, e$_1$ is the error rate of single-photon states. Q$_\mu$ and E$_\mu$ can be measured directly from the experiment, and Q$_1$ and e$_1$ need to be estimated by preset experimental parameters. Here, for each given value of the total loss, we select the optimal values of the intensities $\mu_s$, $\nu_1$ that maximize R$_L$. 

The $Q_1$ and e$_1$ can be estimated by   
\begin{equation}
\begin{split}
    Q_1&\geq\frac{\mu^2e^{-\mu}}{\mu\nu_1-\mu\nu_2-\nu_1^2+\nu_2^2}(Q_{\nu_1}e^{\nu_1}-Q_{\nu_2}e^{\nu_2}\\
    &-\frac{\nu_1^2-\nu_2^2}{\mu^2}(Q_\mu e^\mu-Y_0)),
\end{split}
\label{2}
\end{equation}
\begin{equation}
    e_1\leq\frac{E_{\nu_1}Q_{\nu_1}e^{-\nu_1}-E_{\nu_2}Q_{\nu_1}e^{-\nu_2}}{(\nu_1-\nu_2)Y_1},
\label{3}
\end{equation}
where $\nu_1$ and $\nu_2$ are the number of photons in strong and weak decoy states. Here, we take $\nu_2$ as 0. 

In the presence of green light injection attack, $\mu$ ({$\mu$ $\in$ \{$\mu_s$, $\nu_1$, $\nu_2$\}}) becomes $k\mu$ for a certain k that depends on the attack. 
If Alice and Bob fail to notice the green light injection, here they will estimate the parameters, Q$_1$ and e$_1$, with the observed quantities Q$_{k\mu}$ and E$_{k\mu}$ and the original intensities $\mu$. 
If they are aware of this attack, they estimate the parameters, Q$_1$ and e$_1$, with the changed intensities $k\mu$ and get a true secure key rate. 
In order to correspond to the loss mentioned above, we convert k to $\Delta$Loss. 
For simulation purposes we use the experimental parameters as bellow: background rate, Y$_0$, is 2.6*10$^{-5}$ per pulse; total misalignment error, e$_d$, is 0.01; error correction efficiency, f$_0$, is 1.12; detection efficiency, $\eta_d$, is 0.6. 
% We take \textcolor{red}{$\Delta$ = 1 dB and 3 dB} as the typical value to evaluate the impact of insertion loss changes on the key rate, which matches the variation of insertion loss measured by our experiment. 

\begin{figure*}
\centering 
\includegraphics[width=0.9\linewidth]{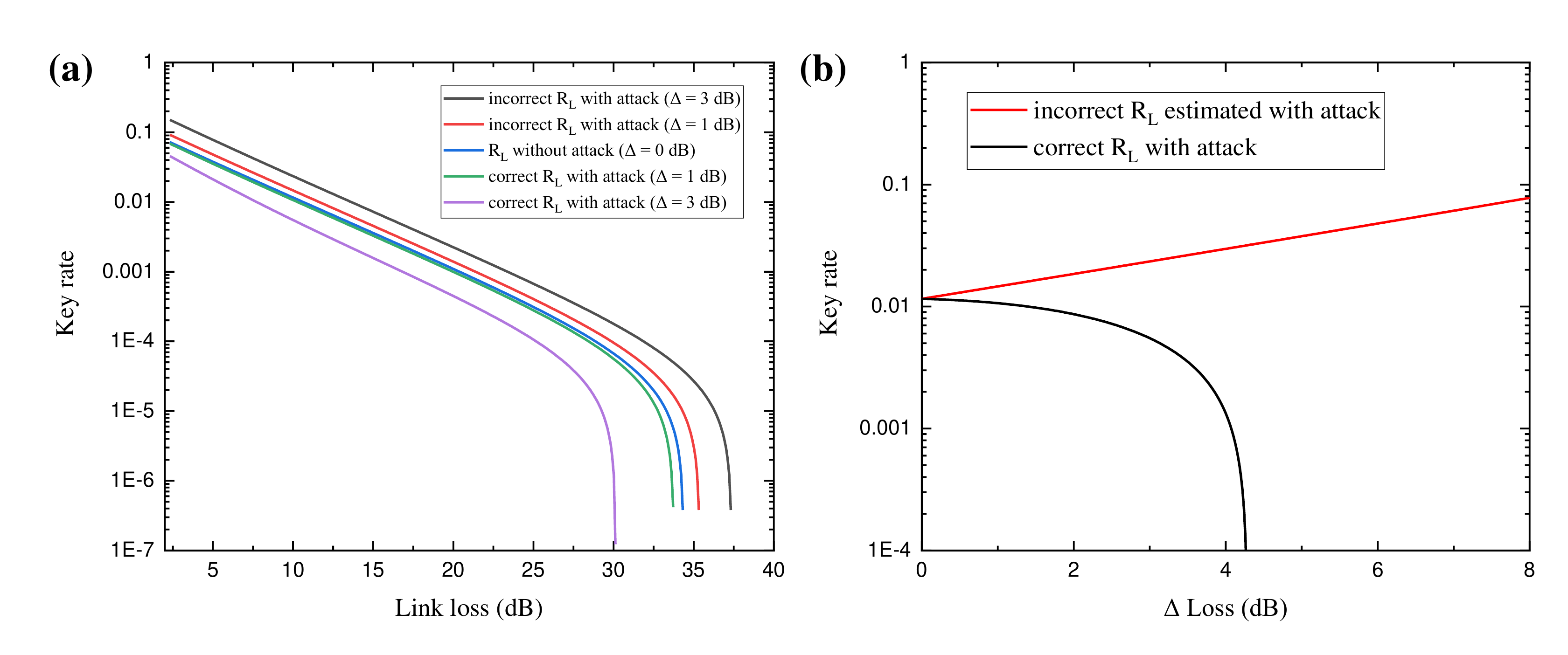}
\caption{(a) Simulated secure key rate as a function of the total loss. 
We take $\Delta$ = 1 dB and 3 dB as the typical values to evaluate the impact of insertion loss changes on the key rate, and the results indicate a significant decrease in the key rate as the insertion loss increases.
In particular, no positive key can be generated when $\Delta$ exceeds 5 dB.
% \textcolor{red}{Moreover, when $\Delta \geq$ 5 dB, the secure $R_{L}$ is no longer positive.}
(b) Lower (R$_L$) bounds on the secret key rate as a function of the parameter $\Delta$Loss for a fixed total loss of 12.22 dB, that contains 10 dB of link loss and 0.6 of detection efficiency.
}
\label{fig6}
\end{figure*}

We take $\Delta$ as different values to evaluate the impact of insertion loss changes on the key rate, and the resulting lower bounds on the secret key rate are shown in Fig. \ref{fig6}. 
The black line indicates the lower bound $R_{L}$ given in the absence of the attack. 
The red solid line shows the value of $R_{L}$ estimated by Alice and Bob if they're not aware of the attack. 
The blue line, on the other hand, illustrates the correct secure value of $R_{L}$ in the presence of an attack.
As we can see in Fig. \ref{fig6}(a), the secure $R_{L}$ is significantly lower than the $R_{L}$ actually estimated by Alice and Bob. 
More precisely, in the presence of the attack, the security proof introduced in Ref. \cite{ma2005} cannot guarantee the security of the key obtained by Alice and Bob. 
Furthermore, as shown in Fig. \ref{fig6}(b), the secure $R_{L}$ drops dramatically as $\Delta$ increases, and it turns out that a larger $\Delta$ may be a compromise of practical QKD. 
In general, legitimate users of the system may significantly overestimate the secure key rate provided by appropriate security proofs in the presence of the attack.

\section{\label{Countermeasures}Countermeasures}
Our experimental results show that none of the $LiNbO_3$-based optical modulators is confidently robust against the green light injection attack.
An eavesdropper can actively control the output photon numbers of the QKD transmitter through external green light injection, which will affect the practical security of QKD systems.
Therefore, countermeasures against such attacks need to be developed. 

According to the test results, modulators from different manufacturers have different performances, and users should carefully choose $LiNbO_3$-based optical modulators, such as those doped with optical-damage-resistant Impurities-MgO. 
Based on the experimental results, it is evident that the device performance varies significantly among different manufacturers and different batches. 
These discrepancies can be attributed to variations in waveguides, crystal orientations, substrate growth, and fabrication processes. Each of these factors influences the response of the samples to externally injected light.
It is important to note that the current fabrication of waveguides is not yet optimized, indicating that there is ample room for improvement. 
This presents an opportunity for the development of more resilient $LiNbO_3$ modulators in the future.
Besides, modulators based on other materials can also be chosen, but similar effects of these modulators should be carefully tested before being applied to QKD systems.

Another option for Alice to protect the QKD transmitter is to apply optical isolators, which can significantly isolate Eve's injection light and make Eve more difficult to attack.
However, the reverse transmission isolation of short wavelengths is rarely studied, and we only note the work of researchers testing the isolation of 1550-nm isolators in the 1500–2100 nm range \cite{nasedkin2022quantum}. 
A wider wavelength range needs to be considered, such as all wavelength bands that the optical fiber can transmit.
We also note another recent work \cite{Tan_2022}, where the reverse transmission isolation of isolators and circulators is significantly reduced under the impact of external magnetic fields. 
This work indicates that the performance of isolators may be compromised when other attacks are launched at the same time.

The third option for Alice to discover eavesdroppers might be to use an incident-light monitor to detect the injected light and attain real-time feedback. 
This includes monitoring both the working wavelength of quantum light and the injected attack light, which typically has a shorter wavelength.
A photodetector (PD) can serve as an effective incident light monitor and can be placed in various configurations, such as after the LD, after the modulators, or after variable optical attenuators \cite{Sibson2017,sax2022highspeed,Minder2019}. 
However, it is important to note that even after a few days, the insertion loss does not fully return to its initial value. 
This poses a challenge for legitimate users to detect Eve without having to keep the external laser continuously active. 
To identify the presence of Eve, Alice would need to diligently record all anomalous events and carefully identify cases that correspond to a light injection attack. 
This task may prove to be challenging. 
Nevertheless, despite the difficulties involved, implementing a monitor is still a feasible approach for detecting eavesdroppers.
% PD can be a good incident light monitor, that might be placed in different configurations after LD, after modulators or variable optical attenuators. \cite{Sibson2017,sax2022highspeed,Minder2019}} 
% However, because the insertion loss does not restore to its initial value even after a few days, it will be more difficult for the legitimate users to detect Eve without keeping the external laser on at all times. 
% In order to discover Eve's existence, Alice needs to record all anomalous events and pick out the cases that belong to green light injection attack, which may be a difficult task. 
% But overall, it's a feasible way to add a monitor.

Another viable solution is to incorporate a narrow-band filter at the transmitting end to effectively mitigate the injection of power by Eve at non-quantum-light working wavelength. 
Various types of filters commonly employed in fiber optic and free-space systems can be utilized for this purpose, such as narrowband filters, Dense Wavelength Division Multiplexing (DWDM) filters, and dichroic filters, among others. 
These filters can effectively block or attenuate undesired wavelengths, ensuring that only the intended quantum signal is transmitted.
% There is no doubt that it would to be a lot harder for Eve to launch an attack if the isolator worked at short wavelength. 
% Additionally, doping $LiNbO_3$ with optical-damage-resistant impurities can lead to a higher photorefractive resistance threshold \cite{kong2012}.

\section{\label{Discussion}Discussion}

In a practical QKD system, there are often more than one $LiNbO_3$ components (e.g. one PM and one IM) in a system. 
These modulators are typically positioned at different locations within the system. 
Consequently, due to the inherent losses of each modulator (typically ranging from 3 to 5 dB) and the impact of other optical components, the actual injected optical power received by each modulator can vary. 
Notably, the outermost modulator tends to receive the maximum injected optical power. 
However, during an actual attack scenario, it is not necessary to specifically target a particular modulator. 
As all modulators in the optical path are impacted by the injected light, the outermost modulator typically receives the highest optical power, while inner modulators receive less. Consequently, all modulators experience increased losses due to the injected light, which accumulate to contribute to the total loss variation.

The light injection attack will inevitably affect the refractive index of the waveguides, consequently inducing a modification in the global phase of the transmitted light \cite{PhysRevApplied.19.054052}. 
This prompts the question of whether such an attack would be effective against a practical QKD system.
In the case of non-phase-encoded QKD systems, such as polarization-encoded systems \cite{Li:19,Agnesi:19,Lucio-Martinez_2009}, where the information is encoded in the polarization state of photons rather than their phase, the laser attack-induced phase change does not result in an alteration of the quantum bit error rate (QBER). 
This raises immediate security concerns for these QKD systems. 
Furthermore, even QKD systems employing phase encoding are not exempt from the threats posed by this attack method.
Practical QKD systems inherently experience instability in their system phase due to factors like variations in ambient temperature along the optical fiber link or atmospheric disturbances. These environmental factors result in gradual drifts in the system's phase. 
To ensure a low QBER, continuous phase feedback is typically employed during experiments \cite{Yuan:05,PhysRevLett.111.130502}.
The attack method proposed in this paper can exploit this vulnerability by executing the attack before the system undergoes phase feedback, making it nearly imperceptible to legitimate users.

In this work, we use 532 nm green light for this light injection attack. 
Ref \cite{luo2012} shows that Zr:Fe:$LiNbO_3$ crystals have larger refractive index change, higher recording sensitivity and larger dual-wave coupling gain coefficient at 473 nm wavelength than at 532 nm wavelength under the same experimental conditions. 
That means that more wavelengths of light can be utilized to carry out such attacks, which makes defenses much more difficult. 
Which wavelength of light works best at the lowest light intensity is also a subject worth studying.
As for free-space QKD, the green light is often used as a beacon light, which will mix and overlap with the green light injected into the attack, making it more difficult to detect.
Because all the modulators exhibit obvious increases in insertion loss, so we accounted for this effect in our security risk evaluation. 

Besides, two of the PM samples show half-wave voltage increases of 1.53 V and 2.81 V, which Eve can also take advantage of.
In the first case, Alice and Bob are unaware of the voltage change in the presence of an attack.
% Incomplete recovery can make the V$\pi$ larger than the initial value. 
The effect of V$\pi$ increase is similar to the phase-remapping attack \cite{Xu_2010}, causing the phase difference between coded states to change. Eve can distinguish between different states by remapping the encoded phase information from \{0, $\pi$/2, $\pi$, 3$\pi$/2 \} to \{0, $\delta$/2, $\delta$, 3$\delta$/2\}, where $\delta$ \textless $\pi$.
The light injection makes the phase-remapping attack not only suitable for bidirectional QKD systems, such as the "plug-and-play" system, but also has a broader range of applications. 
The second scenario involves Alice and Bob calibrating the voltage during the operation of the QKD system. 
This will make $\delta$ \textgreater $\pi$ after the calibration and recovery procedure, which presents another intriguing topic for exploration.
Nevertheless, this phenomenon and the impact of increased half-wave voltage on the security of practical QKD systems can be studied in follow-up works.

\section{\label{Conclusion}Conclusion}
In summary, we experimentally investigate the impact of photorefraction induced by injected green light on $LiNbO_3$-based optical modulators to enhance the security of practical QKD systems. 
We select several $LiNbO_3$ modulator samples to test their properties before and after green light irradiation, and they all exhibit increased insertion loss after green light irradiation. 
At the same time, the half-wave voltage and extinction ratio of the modulators also tend to increase and decrease, respectively.
We also experimentally investigate that all the changes can be recovered by shining weaker green light, providing convincing evidence for a hard-to-find green light injection attack, since Eve can actively control the output photon numbers of the QKD transmitter through external green light injection.
Based on our experimental results, we analyze the security risks of this attack and find that the legitimate users of the system, when attacked, might significantly overestimate the secret key rate provided by proper security proofs. 

Our work highlights the significance of selecting optical modulators, adding optical isolators, and monitoring the injection light power to enhance the security of the practical QKD system.
It is worth noting that the proposed source-side light injection attack is applicable to various types of QKD systems, including not only decoy-state BB84 QKD, but also MDI-QKD and TF-QKD, etc., which can be carried out in the follow-up works.
% A detailed analysis of its effect on decoy-state BB84 and MDI QKD protocols is given in Ref. \cite{huang2019}. 

Note that a related experimental work has been reported in Ref. \cite{PhysRevApplied.19.054052}. 
Both our work and the work described in Ref. \cite{PhysRevApplied.19.054052} exploit the photorefractive effect of $LiNbO_3$-based devices for light-injection attack. 
However, the work described in Ref. \cite{PhysRevApplied.19.054052} changes the attenuation of an MZI-based variable optical attenuator by external light injection, which utilizes a lower injection power.
In contrast, we demonstrate that the corresponding attenuations of two widely-adopted $LiNbO_3$ modulators, i.e. phase modulator and intensity modulator, can both be directly controlled by external light injection with higher power, not limited to MZI-based variable optical attenuators.

% \section{Discussion and outlook}
%In this work, we use 532 nm green light for this light injection attack. 
%Ref \cite{luo2012} shows that Zr:Fe:$LiNbO_3$ crystals have larger refractive index change, higher recording sensitivity and larger dual-wave coupling gain coefficient at 473 nm wavelength than at 532 nm wavelength under the same experimental conditions. 

\begin{acknowledgments}
The authors thank Anqi Huang and Vadim Makarov for their helpful discussions. This work was supported by National Key Research and Development Program of China (2020YFA0309701, 2020YFA0309703, 2020YFE0200600, 2022YFF0610100), Shanghai Municipal Science and Technology Major Project under Grant 2019SHZDZX01 and Anhui Initiative in Quantum Information Technologies.
% Thanks are due to Prof. Jie Liu and the Single Event Effect Test Beamline Terminal of the Heavy Ion Research Facility in Lanzhou for assisting with the irradiation detection in this study.
\end{acknowledgments}

\appendix
\section{\label{appendix A}Stability of the insertion loss}
\begin{figure*}
\centering 
\includegraphics[width=0.9\linewidth]{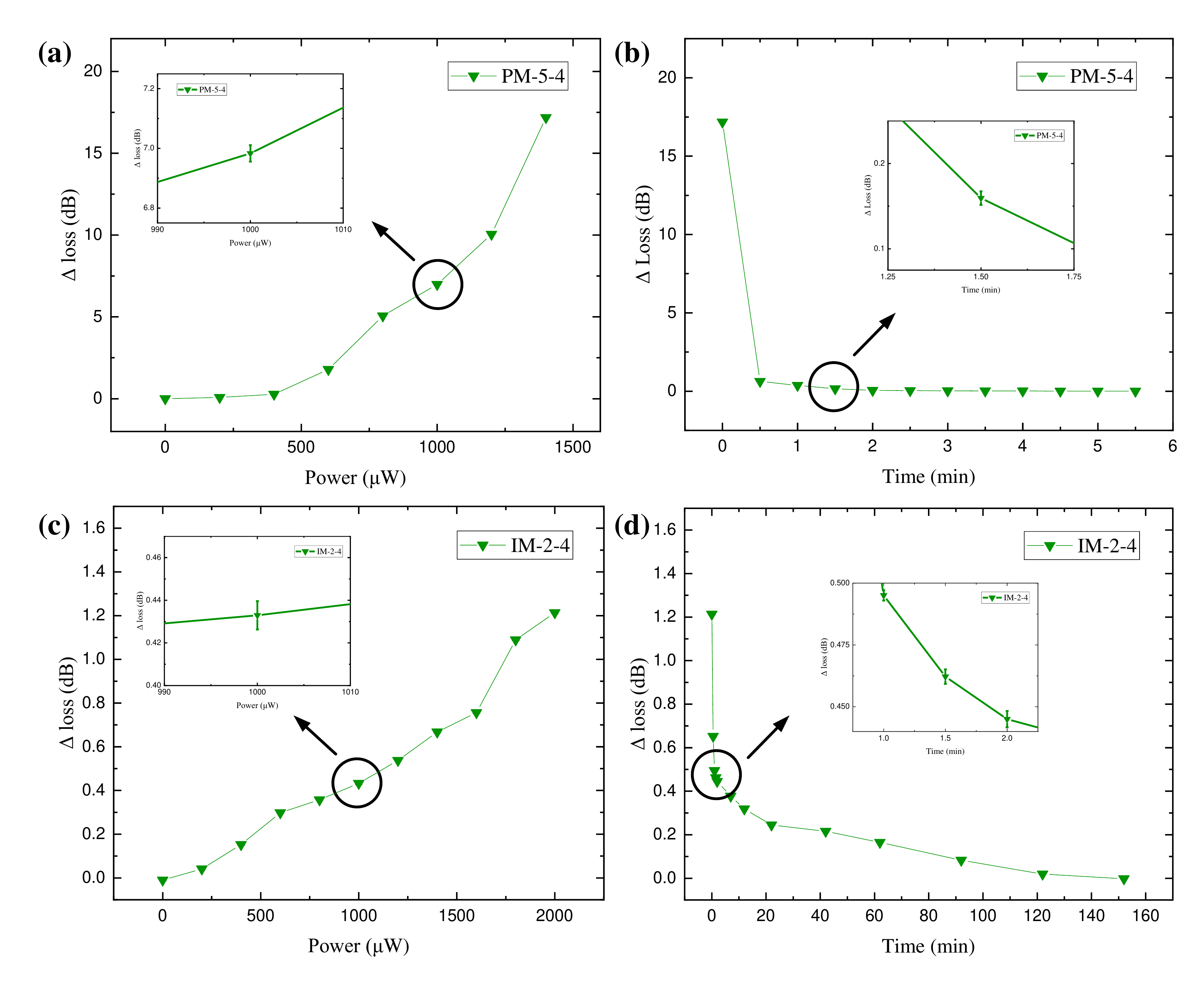}
\caption{
(a) Stability of increased insertion loss of PM-5. 
We increase the power of the incident green light to 1.4 mW with a step size of 200 $\mu$W, with an exposure time of 5 minutes per step. 
Then we record the measured insertion loss continuously for a duration of 5 minutes, and the observed fluctuation remains within in a range of less than 0.1 dB. 
(b) Stability of recovered insertion loss of PM-5.
The observed fluctuation remains within in a range of less than 0.1 dB. 
(c) Stability of increased insertion loss of IM-2.
We scan the applied voltage 10 times in 5 minutes and obtain ten sets of insertion loss to calculate the stability. 
The observed fluctuation remains within in a range of less than 0.02 dB. 
(d) Stability of recovered insertion loss of IM-2. 
The observed fluctuation remains within in a range of less than 0.02 dB.
}
\label{fig7}
\end{figure*}

We have added the stability measurements of the insertion loss of PM-5 and IM-2, and the results are shown in Fig. \ref{fig7}.
Due to the minimal magnitude of the error bars, we have zoomed in on specific points in the subplots to display the values accurately, indicating that they are consistently less than 0.1 dB and 0.02 dB, respectively. 
This observation confirms the stability of the insertion loss test.
% In Sec. \ref{exp_setup}, we mainly show the measurement results of change in insertion loss of 5 PMs and 2 IMs under the illumination of green light. 
% In order to show the stability of the insertion loss after each exposure,  we choose PM-5 and IM-2 to record the statistical fluctuations of insertion loss after each exposure. 
% The testing procedure differs little from that described in Sec. \ref{exp_setup}, specifically we add 5 minutes of continuous insertion loss test after each green light exposure. 
% As for PM-5, we record the insertion loss once per second for 5 minutes after each green light exposure. 
% As for IM-2, since there is no way to record the maximum and minimum interference values at the same time, we scan the voltage 10 times in 5 minutes and obtain 10 sets of insertion loss to calculate the error bar. 
% The results are shown in Fig. \ref{fig7}. Due to the errror bar is too small, we have enlarged some points in the subgraph to show the value of the errror bar, that is less than 0.03 dB. It is proved that photorefractive effcct can also exist stably in the absence of external light.

\section{\label{appendix B}A Comparison between forward injection test and reverse injection test}
\begin{figure}
\centering 
\includegraphics[width=0.9\linewidth]{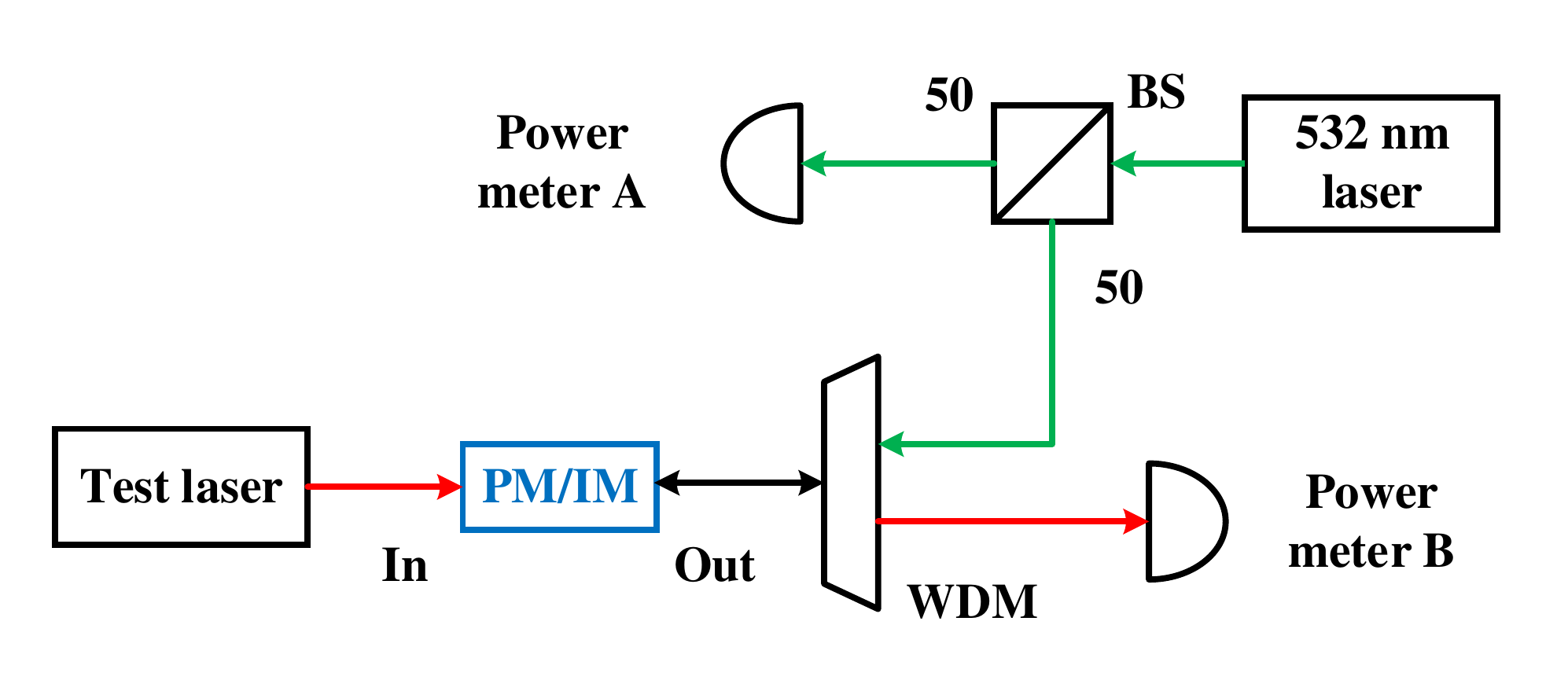}
\caption{
Simplified diagram of experimental setup of insertion loss, where green light entering from the reverse direction.
}
\label{fig8}
\end{figure}

\begin{figure*}
\centering 
\includegraphics[width=0.9\linewidth]{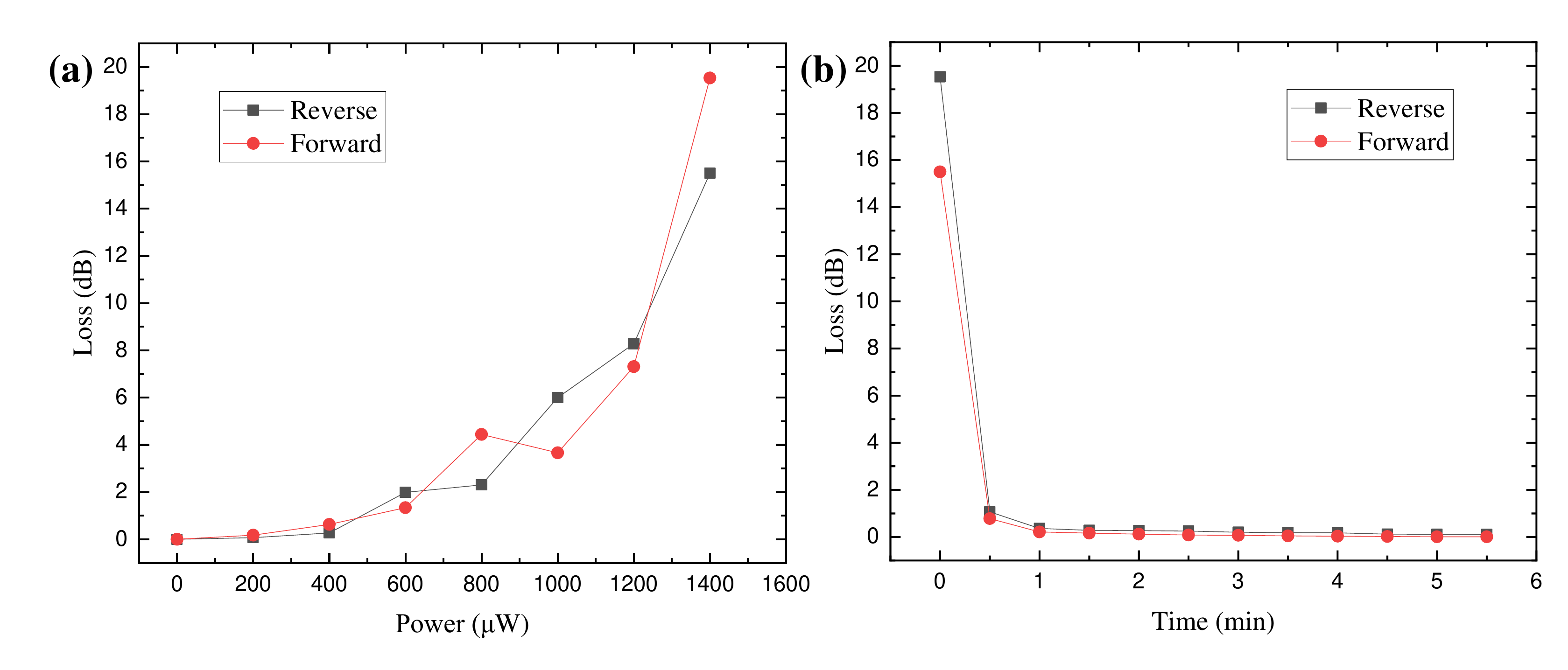}
\caption{
(a) Test results of increased insertion loss of PM-5 in the case of forward light injection and reverse light injection. 
(b) Test results of recovered insertion loss of PM-5 in the case of forward light injection and reverse light injection.
The black curves indicate the test results of the insertion loss and recovered insertion loss where injection light enters from the forward direction, which has been proposed in Fig. \ref{fig3}.
The red curves indicate the results of supplementary tests with reverse light injection direction. 
}
\label{fig9}
\end{figure*}

% Additional descriptions of the attack and testing methods}
It should be noted that the direction of light injection in the experimental setup (Fig. \ref{fig2}) is opposite to that in the proposed light injection attack (Fig. \ref{fig5}).
Here, we conducted a supplementary test with reverse light injection direction (Fig. \ref{fig8}), and the results are directly compared with previous results in the case of forward light injection (Fig. \ref{fig9}).
Despite minor variations between the two tests, the overall trend of the test results remains consistent. 
This suggests that different light injection directions have no significant impact on the insertion loss of the tested modulators.

% As shown in in Fig. \ref{fig6}, the direction of eve's attack light is the opposite of the direction of the light Alice uses to measure insertion loss. In Sec. \ref{exp_setup}, we use the 1550 nm light in the same direction as eve's 532 nm laser to test the insertion loss of modulators, that is slightly different from the actual execution of the attack. We conduct another test process for PM-5 as shown in Fig. \ref{fig8}.We drive green light from the output of PM-5 and the test the insertion loss using the test laser from the input to output, that is consistent with the actual attack. The results are shown in Fig. \ref{fig9}.} 

% \textcolor{red}{
% We compare the measured results with those in Sec. \ref{exp_setup}, and find that there is the same trend in the change of insertion loss. Note that the test result of synthetic insertion loss is also applicable, though reverse insertion loss is used in actual attack.} 

\end{document}